# MANAGING CELLULAR BILLING PLAN SWITCHINGS

*V.Ya. Vilisov*
*University of Technology, Russia, Moscow Region, Korolev*
*vvib@yandex.ru*

***Abstract.*** *Here we shall consider a very popular practical applied problem of managing mode switching (in this work we are considering managing billing plans). Out of the two parties (service provider and service consumer), participating in the processes modelled here, we shall consider only a consumer type of a problem. Herein we provide formal characterization of the problem as well as the elements necessary for its solution. We shall consider full predicted costs, originating when switching to a billing plan as a target index. The work contains an example that provides a detailed view of the application technology referring to the suggested problem solution algorithm. Using the example's data we have performed the analysis measuring the problem's sensitivity in relation to the growth of the traffic volume. Herein we provided a polynomial approximation of the target index value depending on the traffic volume.*

## *1. Introduction*

Different activity spheres very often possess the possibility for switching between the modes, variants, schemes, tariffs and so forth [4-7]. For example:
- Enterprises can periodically
  - Switch between different providers of raw materials;
  - Switch between different taxation modes (general, simplified etc.);
  - Use, for the purposes of the works and/or outsourcing services performance, some external company (for the purposes of providing communal, bank, tourist, legal and other services) or to execute the works by its own efforts;
- Natural persons can do regular shopping in different supermarkets or to make orders via Internet;
- Consumers of mobile communication service (or Internet services) can periodically switch between different billing plans. The reason of the latter can lie in the changes of the billing plan multitude that is offered by the provider to its consumers or in the change of the consumer's needs regarding the traffic's volume and structure.

The main elements of the switching managing problem are similar for different applications, therefore let us consider switching modelling using the example of mobile communication services. Here the billing plans (BP) constitute the work modes between which the switching takes place, while the task of the switching managing lies in the periodic selection of the most beneficial mode.

The selection of a variant (mode) depends on the characteristics of both the service (products) range, offered by the provider and the consumer's own needs. Therefore, it is generally possible to consider both the task for the provider to choose the optimal mode range (e.g. the range of the billing plans) for the multitude of consumers and the task for the consumer to choose the most beneficial mode (according to his/her needs and target preferences). Herein we shall consider only the consumer's selection task. At that we think that the dynamic pattern for the mode range and mode parameters is significantly lower that the consumer's response to these changes and to the variations of his/her own needs.

## *2. Billing Plan Switching Management Problem Set-Up*

The optimal BP selection scheme is provided on Picture 1.

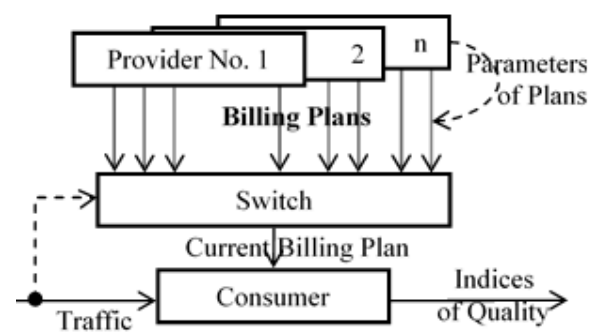

*Pic. 1. Billing Plan Management Scheme*

The task: using the predicted traffic of future planning period to choose the best BP out of the existing ones at the time of selection.

Without losing the communality, let us decide that the traffic parameters are the same as they were during the previous period.

The task consists of two main elements:
1. Traffic (with corresponding characteristics).
2. Billing plans (with their own parameters).

Main index shall be the sum of the payment related to the monthly services, provided according to the tariffs of some BP. Planning interval shall be one month, because providers allow switching only starting from the first day of each month.

Criterion is the minimum monthly payment amount.

Solution shall be a number of the billing plan that provides the performance of the criterion.

Let us consider main properties of the traffic and the BP.

*2.1. TRAFFIC*

The traffic, being the chronological printout of the services, provided to the consumer in accordance with the capabilities of the current BP (and the services, included by the user), includes: service type, volumes, costs and other characteristics. The extract from such printout is provided in Table 1.

*Table 1. Billing Plan Data Detail Extract*

| Date | Time | Number | Telecommunication Enterprise Zone | Service | Duration, min:sec | Total without VAT |
|---|---|---|---|---|---|---|
| 19.08.2010 | 14:29:27 | +7495… | Moscow | Tel | 2:23 | 7.627 |
| 20.08.2010 | 12:01:27 | +7985… | Moscow | Tel | 0:57 | 2.542 |
| 20.08.2010 | 12:14:39 | +7910… |  | SMS | 1 | 4.449 |
| 20.08.2010 | 12:15:55 | +7916… | Moscow | Tel | 0:33 | 2.542 |
| 24.08.2010 | 17:29:04 | +7926… |  | SMS | 1 | 1.652 |
| 25.08.2010 | 10:02:17 | +7495… | Moscow | Tel | 0:44 | 2.542 |

Let us provide main *prerequisites and delimitations* for the billing plans that are reviewed in the task.

Variety of products that are provided to consumers today is quite big, but the leading products here include:
- Telephone calls (hereinafter referred to as *calls*);
- *SMS*;
- Internet.

Let us consider the problem solution technology using the call traffic (only outgoing calls). Inclusion of other components of the full traffic can be performed in the same way as the inclusion of calls, with the value of target integrated index in relation to all components of the full traffic be calculated by summing the costs of all components.

The most important traffic parameters are the following (see Pic. 2):
- Duration ($\theta$) of every call (in minutes).
- Time intervals between the calls ($\tau$) that are measured in minutes.

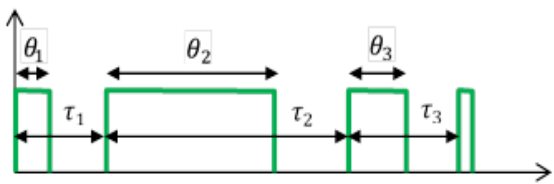

*Pic. 2. Traffic Parameters*

These parameters have a random nature. For example, the $\theta$ bar chart can look like Picture 3.

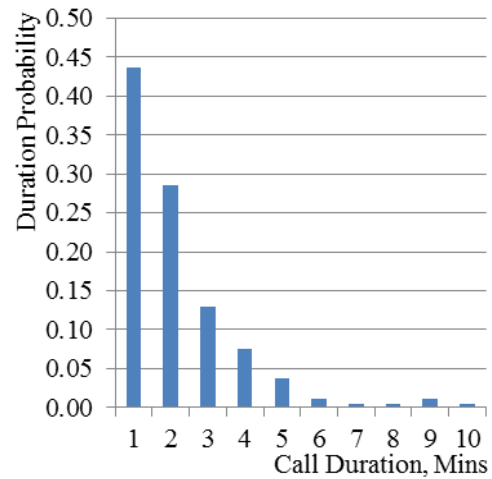

*Pic. 3. Call Duration Bar Chart*

I.e. the distribution density can be analytically represented using the exponential way [3]:

$$f(\theta) = \mu e^{-\mu\theta}. \qquad (1)$$

Such way is convenient for the analysis because it is defined by the only parameter $\mu$. Still, it is not necessary to apply analytical representation in order to solve the BP management problem.

Periodicity (intervals between the calls) is usually represented as an exponential distribution:

$$f(\tau) = \lambda e^{-\lambda \tau}. \qquad (2)$$

Knowledge of $\lambda$ and $\mu$ traffic parameters (or their sample estimates, obtained from the bar charts) allows solving the best BP selection problem either analytically [1] or using the means of simulation modelling, i.e. by "pushing" the traffic through each one of the alternative BPs, calculating the costs and choosing the cheapest one (like in [4]).

*2.2. BILLING PLANS*

Billing plans can be characterized by the multitude of technical parameters that correspond to the different prices of the plan's separate products like:
- Limitation parameters (limited, unlimited);
- Area (Moscow and oblast, other regions of Russia, foreign countries);
- Communicant's belonging to the BP (similar BP, similar network, other mobile networks, urban network and others) etc.

Basing on these parameters, call traffic can be divided into smaller subgroups that correspond to different conditions.

When evaluating BP it is necessary to consider fixed costs connected with the change of a BP or with the change of a provider.

*2.3. THE BEST BILLING PLAN SELECTION ALGORITHM*

If to consider the subscriber's full traffic that consists of separate components $k = 1, \ldots, K$ (calls, SMS, Internet etc.), the cost of the traffic at the time of its realization by $i$-numbered BP ($i = 1, \ldots, M$) shall be the objective function (OF) of the selection problem:

$$L_i = \sum_{k=1}^{K} L_i^k, \qquad (3)$$

where the number of BP shall be its argument. At that the selection criterion is:

$$i_{opt} = \arg \min_{i=1,\ldots,M} L_i. \qquad (4)$$

In order to simplify the calculations we shall consider *only the call traffic* ($k = 1$), i.e. we shall omit superscript during the further explanation.

Usually, for the purposes of some conditions, each BP is set with the price of one minute talk (payoff function $v(\theta)$). Thus, $v(\theta)$ is the random argument function. Random argument distribution (call duration $\theta$) is defined by the density $f(\theta)$ (e.g. in the form (1)).

Average value of one random call cost shall be defined using the standard formula for the random argument function's mathematical expectation [3]:

$$s = \int_0^\infty v(\theta) f(\theta) d\theta. \qquad (5)$$

$\lambda$ index represents the call traffic, i.e. the average number of calls per some unit of time (e.g. $\lambda = 210$ 1/month). Thus, for $i$-numbered BP the target function shall look like:

$$L_i = \lambda s_i = \lambda \int_0^\infty v_i(\theta) f(\theta) d\theta. \qquad (6)$$

Let us consider how we can calculate the subgroups that are charged differently within the call traffic (for example, the calls to the subscribers of the same BP as the one under analysis or the calls to the landline subscribers etc.).

These subsets shall be defined by $j = 1, \ldots, N$ index. Number of these subsets ($N$) can vary for different BPs, but we shall consider them similar for the simplicity of formal characterization, while the excessive ones shall be considered empty.

Every BP has its own elements for each $j$-numbered subflow: payoff function $v_{ij}(\theta)$; call traffic distribution $f_j(\tau)$ with $\lambda_j$ parameter; distribution of their durations $f_j(\theta)$ with the parameter $\mu_j$.

Thus, for BP$_i$ the OF (6) shall look like:

$$L_i = \sum_{j=1}^{N} \lambda_j s_{ij} = \sum_{j=1}^{N} \lambda_j \int_0^\infty v_{ij}(\theta) f_j(\theta) d\theta. \qquad (7)$$

Here, $s_{ij}$ is the average cost of one call for $j$-numbered subgroup of $i$-numbered BP.

For the purposes of the discrete distribution of the durations ($\theta = 1, 2, \ldots, T$ minutes), the expression (7) can be represented as the sum:

$$L_i = \sum_{j=1}^{N} \lambda_j s_{ij} = \sum_{j=1}^{N} \lambda_j \sum_{\theta=1}^{T} v_{ij}(\theta) f_j(\theta). \qquad (8)$$

At that the payments can be shown as Table 2.

Usually, $v_{ij}(\theta)$ function is a graduated (piecewise constant) function of $\theta$ argument per each $i$-numbered BP within the $j$-numbered conditional group that corresponds to it.

The calculation algorithm includes the following.

1. To analyze all BPs that can be chosen for switching by selecting the conditional groups and subgroups, constituting the whole traffic.
2. To define the payoff functions $v_{ij}(\theta)$ per each subgroup of every BP.
3. To divide statistic data of the traffic (retrospective or predicted) into the subgroups similar to the ones, provided in Table 1, to calculate the estimates $\lambda_j$ and $\mu_j$ for each of them or to use the data as empirical distribution (bar charts).

4. To calculate the values of one call's average cost for $j$-numbered subgroup of $i$-numbered BP ($s_{ij}$), and, taking into consideration the forces of $\lambda_j$ as the weights, to obtain the values of $L_i$ for all BPs (see Table 2). $L_i$ values are the variable expenses (depending on $\lambda_j$ and $\mu_j$) for the corresponding BPs, but there also exist fixed expenses.

5. To estimate the fixed expenses $L_i^0$ for each $i$-numbered BP for the forthcoming (planned) month. Fixed expenses depend on the current BP and current provider because only this data defines the cost of the provider switching or switching the BP within the product range of the current provider. Moreover, prehistory of BP switching is also important, as it may occur that the decision taker has an active *SIM*-card, which could be used by only paying the other BP switching fee without buying it. Subscription fee is also a part of the fixed costs.

*Table 2. Variable Expenses for the Conditional Subgroups*

| Billing Plans | Conditional Subgroups | | | | | Average Monthly Payment Amount |
|---|---|---|---|---|---|---|
| | 1 | 2 | ... | $j$ | ... | $N$ | |
| | $\lambda_1$ | $\lambda_2$ | ... | $\lambda_j$ | ... | $\lambda_N$ | |
| BP$_1$ | $s_{11}$ | $s_{12}$ | ... | $s_{1j}$ | ... | $s_{1N}$ | $L_1$ |
| BP$_2$ | $s_{21}$ | $s_{22}$ | ... | $s_{2j}$ | ... | $s_{2N}$ | $L_2$ |
| ... | ... | ... | ... | ... | ... | ... | ... |
| BP$_i$ | $s_{i1}$ | $s_{i2}$ | ... | $s_{ij}$ | ... | $s_{iN}$ | $L_i$ |
| ... | ... | ... | ... | ... | ... | ... | ... |
| BP$_M$ | $s_{M1}$ | $s_{M2}$ | ... | $s_{Mj}$ | ... | $s_M$ | $L_M$ |

Taking into account full costs ($L^F = L_i + L_i^0$) when switching from the current BP to another $i$-numbered BP, it is important to choose the one that provides the minimum of the full costs.

## 3. Switching-Optimal Billing Plan Selection Example

Let us use the data (service data detail) that was received for the 6-month period by the subscriber of the cellular provider MTS with the current BP being "Oblastnoi". As alternative BPs, without losing communality and for the purposes of better result-readability, let us consider only currently active five BPs offered by MTS (as of spring 2011).

Calculation of the average value $m_\theta$ and root-mean-square deviation (RMSD) $\sigma_\theta$ shows that they are quite close (about 2.45), therefore the sampling distribution can be approximated by the exponential $f(\theta) = \mu e^{-\mu\theta}$ with $\mu = \frac{1}{m_\theta} = \frac{1}{\sigma_\theta} = 0{,}41$ parameter.

In order to calculate the variable expenses we shall use form (8). The extracts of the payoff functions of all five BPs under review are provided in Tables 3-7. The BP, provided in Table 8 is the current one but MTS does not offer it for sale (it is not active as it is not included into the price-list) therefore it is not possible to switch to it from another BP. Still, it is possible to leave it should it be more beneficial than others.

Concrete value of fixed costs for a BP depends on the following data: what BP was the current one, does the subscriber have a SIM-card of this BP's provider and what provider currently is his/her current one. Let us define the constant component of BP's cost index as $L_i^{constant}$.

The forces ($\lambda_{ij}$) of traffic (average number of the monthly calls for the period under review) per each $j$-numbered conditional subgroup for each $i$-numbered BP is provided in Table 10. The variable costs, represented by the objective function (8) shall look like (9).

$$L_i = \sum_{j=1}^{N} \lambda_{ij} s_{ij} = \sum_{j=1}^{N} \lambda_{ij} \sum_{\theta=1}^{T} v_{ij}(\theta) f_j(\theta). \quad (9)$$

Fixed costs of all BPs are provided in Table 9.

One-call average expenses per each $j$-numbered conditional subgroup (per each $i$-numbered BP) are shown in Table 11 (analogue of Table 2).

*Table 3. BP₁ "Super Zero"*

| Mins | Subgroups | |
|---|---|---|
| | To MTS Numbers | To Other Numbers |
| 1 | 2,5 | 3,5 |
| 2 | 0 | 0 |
| 3 | 0 | 0 |
| 4 | 0 | 0 |
| 5 | 0 | 0 |
| 6 | 2,5 | 3,5 |
| 7 | 2,5 | 3,5 |
| … | 2,5 | 3,5 |
| 13 | 2,5 | 3,5 |
| 14 | 2,5 | 3,5 |
| 15 | 2,5 | 3,5 |
| 16 | 2,5 | 3,5 |
| … | 2,5 | 3,5 |

*Table 4. BP₂ "Maxi Plus"*

| Mins | Subgroups |
|---|---|
| | To All Numbers |
| 1 | 0 |
| 2 | 0 |
| 3 | 0 |
| 4 | 0 |
| 5 | 0 |
| 6 | 0 |
| 7 | 0 |
| … | 0 |
| 149 | 0 |
| 150 | 0 |
| 151 | 2,2 |
| 152 | 2,2 |
| … | 2,2 |

*Table 5. BP₃ "Many Calls to All Networks"*

| Mins | Subgroups |
|---|---|
| | To All Numbers |
| 1 | 3,5 |
| 2 | 3,5 |
| 3 | 3,5 |
| 4 | 3,5 |
| 5 | 3,5 |
| 6 | 0,05 |
| 7 | 0,05 |
| … | 0,05 |
| 29 | 0,05 |
| 30 | 0,05 |
| 31 | 3,5 |
| 32 | 3,5 |
| … | 3,5 |

*Table 6. BP₄ "Red Energy"*

| Mins | Subgroups | |
|---|---|---|
| | To Mobiles | To Landlines |
| 1 | 2,2 | 4,2 |
| 2 | 2,2 | 4,2 |
| 3 | 2,2 | 4,2 |
| 4 | 2,2 | 4,2 |
| 5 | 2,2 | 4,2 |
| 6 | 2,2 | 4,2 |
| 7 | 2,2 | 4,2 |
| … | 2,2 | 4,2 |

*Table 7. BP₅ "Ultra"*

| Mins | Subgroups | |
|---|---|---|
| | To MTS | To Other Numbers |
| 1 | 0 | 0 |
| 2 | 0 | 0 |
| … | 0 | 0 |
| 3699 | 0 | 0 |
| 3700 | 0 | 0 |
| 3701 | 0 | 2 |
| 3702 | 0 | 2 |
| … | 0 | 2 |

*Table 8. BP₆ "Oblastnoi"*

| Mins | Subgroups | |
|---|---|---|
| | On Work Days | On Weekends |
| 1 | 3 | 1 |
| 2 | 3 | 1 |
| 3 | 3 | 1 |
| 4 | 3 | 1 |
| 5 | 3 | 1 |
| 6 | 3 | 1 |
| 7 | 3 | 1 |
| … | 3 | 1 |

*Table 9. Components of BP's Fixed Expenses*

| No. of BP ($i$) | Billing Plan | Subscription Fee, Rubles | BP Switching Cost, Rubles | BP Buying Cost, Rubles |
|---|---|---|---|---|
| 1 | BP₁ "Super Zero" | 0 | 90 | 195 |
| 2 | BP₂ "Maxi Plus" | 225 | 90 | 150 |
| 3 | BP₃ "Many Calls to All Networks" | 0 | 90 | 195 |
| 4 | BP₄ "Red Energy" | 0 | 250 | 195 |
| 5 | BP₅ "Ultra" | 2500 | 250 | 1300 |
| **6** | **BP₆ "Oblastnoi"** | **0** | **-** | **-** |

*Table 10. Call Traffic In Relation to Conditional Subgroups and Billing Plans*

| No. of BP ($i$) | Conditional Subgroups ($j$) | | | | | | | Total |
|---|---|---|---|---|---|---|---|---|
| | To All Numbers | To MTS Numbers | To Other Numbers | To Mobiles | To Landlines | On Work Days | On Weekends | |
| | 1 | 2 | 3 | 4 | 5 | 6 | 7 | |
| | $\lambda_{i1}$ | $\lambda_{i2}$ | $\lambda_{i3}$ | $\lambda_{i4}$ | $\lambda_{i5}$ | $\lambda_{i6}$ | $\lambda_{i7}$ | |
| 1 | 0 | 23 | 16 | 0 | 0 | 0 | 0 | 39 |
| 2 | 39 | 0 | 0 | 0 | 0 | 0 | 0 | 39 |
| 3 | 39 | 0 | 0 | 0 | 0 | 0 | 0 | 39 |
| 4 | 0 | 0 | 0 | 30 | 9 | 0 | 0 | 39 |
| 5 | 0 | 23 | 16 | 0 | 0 | 0 | 0 | 39 |
| **6** | **0** | **0** | **0** | **0** | **0** | **33** | **6** | **39** |

*Table 11. Subgroup and Billing Plan Expenses*

| No. of BP ($i$) | One-Call Average Cost Per Subgroup $s_{ij}$ | | | | | | | Total for Variable Expenses, Rubles | Full Costs, Rubles |
|---|---|---|---|---|---|---|---|---|---|
| | To All Numbers $s_{i1}$ | To MTS Numbers $s_{i2}$ | To Other $s_{i3}$ | To Mobiles $s_{i4}$ | To Landlines $s_{i5}$ | On Work Days $s_{i6}$ | On Weekends $s_{i7}$ | | |
| 1 | | 1,16 | 1,63 | | | | | 53 | 143 |
| 2 | 0 | | | | | | | 0 | 315 |
| 3 | 3,13 | | | | | | | 122 | 212 |
| 4 | | | | 2,18 | 4,16 | | | 103 | 353 |
| 5 | | 0 | 0 | | | | | 0 | 2750 |
| 6 | | | | | | 2,97 | 0,99 | 104 | 104 |

In Table 11 one-call average cost was calculated as:

$$s_{ij} = \sum_{\theta=1}^{T} v_{ij}(\theta) f_j(\theta), \quad (10)$$

where payoff function $v_{ij}(\theta)$ per each $i$-numbered BP is set in Tables 3-8, and the discrete distribution of probabilities $f_j(\theta)$ is the bar chart (see Pic. 3) containing per-minute breakdown with $\theta$ being the minute number.

Variable expenses $s_{ij}$ were calculated per each $j$-numbered subgroup of every $i$-numbered BP using formula (10). Total variable expenses were calculated afterwards (see column "Total Variable Expenses" in Table 11) of each BP:

$$L_i^{variable} = \sum_{j=1}^{N} \lambda_{ij} s_{ij}. \quad (11)$$

"Full Costs" column in Table 11 is formed as the sum of variable and fixed costs per each BP:

$$L_i = L_i^{variable} + L_i^{constant}. \quad (12)$$

According to the BP indices shown in "Full Costs" column it is clear that the current $BP_6$ variant is the optimal one both for the traffic parameters under consideration and for the provided BP multitude:

$$i_{opt} = \arg \min_{i=1,\ldots,6} L_i = 6. \quad (13)$$

When switching, the full cost (average full cost of the following month) BP ranking is the following: $R = \{6,1,3,2,4,5\}$. Therefore, the best variant is the current BP No. 6 (see Pic. 4) that minimizes the following month full costs on condition that traffic will stay the same.

Thus, the problem is solved for the current traffic. Still, we have not yet said whether this solution is optimal for the situations with the traffic deviations occurring within the following planned period. In order to answer this question we can conduct the research measuring the solution's sensitivity in relation to the changes of the traffic parameters, e.g. to its total volume. Still, for the purposes of briefness let us suggest that the proportions found between the traffic's separate subgroups shall stay unchanged so that we could compare the variants using the proportional changes of values contained in Table 10.

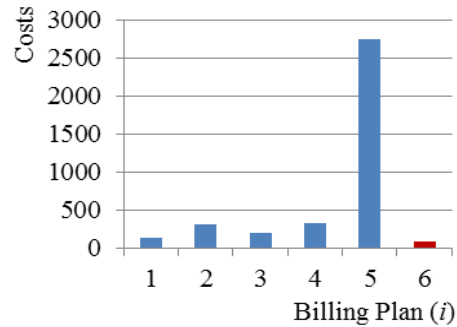

*Pic. 4. The Costs Originating When Switching to Alternative BPs*

When multiplying the total volume of the traffic by $k$ times, other variants of billing plans also become optimal (see Pic. 5).

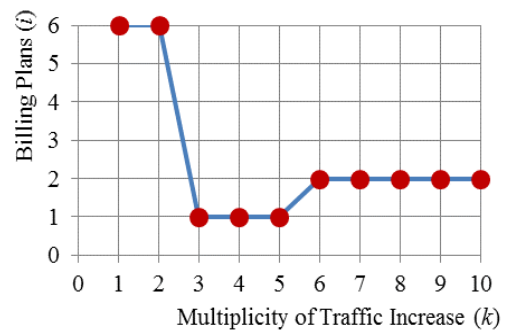

*Pic. 5. Sensitivity of the Problem Solution*

Picture 6 contains changes of the communication service full costs ($L$) that are dependent on the multiplicity of the traffic increase ($k$). The square marker chart shows the traffic cost changes should there occur any changes in the traffic increase and should we not manage the billing plan switching by staying in the original BP$_6$. The regression equation, built in this case with Regression Analysis option of Data Analysis, the MS Excel's add-in, is represented as (14).

$$L = 80k. \qquad (14)$$

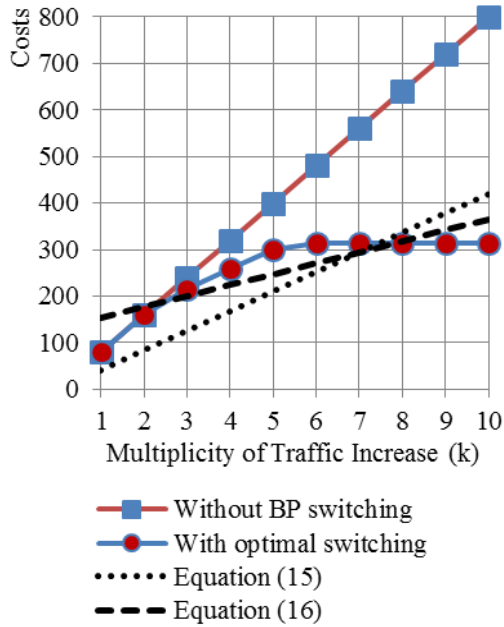

*Pic. 6. Monthly Traffic Full Cost with/without BP Switching (provided for different multiplicities of its growth)*

Round marker chart shows the cost dependence taking into consideration the switching. When using the computational points for building the linear regression equation not containing the constant component, we shall get:

$$L = 42,039k. \qquad (15)$$

It shows that should the traffic total volume increase and should we optimally manage the switching, the average cost of the BP will be twice as less as without the switching. Still, regression model (15) possesses a low value of the concordance coefficient ($R^2 = 0,157$), therefore its adequacy degree is not sufficient for drawing out the conclusions. Construction of the linear model containing a constant component gives us the following result:

$$L = 129,67 + 23,52k, \qquad (16)$$

where we receive a much better concordance coefficient ($R^2 = 0,746$) and the quadratic model leads to a sufficient adequacy degree of the regression model ($R^2 = 0,987$):

$$L = 13,5 + 81,60k - 5,28k^2. \qquad (17)$$

It is possible to get a higher accuracy by building the regression equation of the third degree ($R^2 = 0,997$):

$$L = 24,83 + 115,60k - 12,65k^2 + 0,45k^3. \qquad (18)$$

Thus, regression models (17) or (18) possess an acceptable adequacy degree and can be used for predicting the cellular communication costs should the total volume of the traffic increase (without changing its internal structure).

## 4. Conclusions and Discussion

Usage of the constructed regression models allows predicting the cost of the cellular communication should the total volume of the traffic change (while its internal structure stays the same) without conducting all those detailed calculations provided above. Still, these calculations should be conducted for the given repertoire of the alternative billing plans and traffic only when there is a change in the range of the alternative BPs and/or when the traffic structure changes.

For the purposes of estimating the impact of other traffic parameters upon the cellular communication costs it is possible to build a multidimensional regression model using the technology, provided above.

Lately some cellular providers published simplified calculators on their websites with the purpose being to help choosing their best BP according to the parameters of the expected traffic set by the user. At that such calculators do not take into consideration BPs of other providers. Moreover, there also exist doubts regarding the credibility of such calculations (they could be promoting some new BPs, which are beneficial for the provider).

There is also a software of the independent developers, for example Tarifer [4], which analyzes the all-provider BP range. Still, the Tarifer's working algorithm is a total mystery, while the technology explained in this work is very clear, allows conducting a parameter analysis of the solution, and, when it is necessary, it provides the set level of the cost's probability belief.

The website [4] provides a curious statistics proclaiming that all their visitors who have used Tarifer software for the purposes of choosing the optimal BP (and this information is credible as the analysis was conducted by the website itself) saved around 40% of their cellular costs (we think that the numbers are understated). These estimates are based on the sample that includes more than 30000 software boots. Still, if we imagine that all-country cellular users shall decide to switch to the optimal BPs, we can expect that the profit of the cellular providers will drop accordingly. But, and here we take the public reports of Vimpelcom JSC as an example, its operating activity for the first quarter of 2012 constitutes only 23%. In this situation we cannot but think of an analogy with the column of soldiers lockstepping on the bridge, which can resonate and finally collapse. However, this will not happen to Vimpelcom JSC, because despite our obsession with altruism, we do not want the bridge to collapse underneath our feet, also not desiring to refute the opinion of H. A. Simon [2] who observed the property of bounded rationality inherent to many people. In this case this property can be understood in such a manner that the providers offer us their services with the price being 40% less than the price at which we want to buy them and at which we buy them finally...

Presence of this significant volume of funds (40%), which are easily spent by the clients of the cellular providers caused the appearance of the trust service companies managing billing plans of the clientele. However, even if the final cellular users finally "see the light" and start saving this 40% by themselves or with the aid of trust companies, we think that the self-preservation instinct will make the providers to change the rules of the game. And in this case we will have to conduct additional research and to build other models that would analyze and explain the rational behavior of the subscribers.